\documentclass[aip,preprint]{revtex4-1}

\newcommand{\Tc}{T$_{\rm c}$}
\newcommand{\ca}{$c/a$}
\newcommand{\CFS}{Cs$_x$Fe$_{2-y}$Se$_2$}
\newcommand{\RFS}{Rb$_x$Fe$_{2-y}$Se$_2$}
\newcommand{\KFS}{K$_x$Fe$_{2-y}$Se$_2$}

\usepackage{graphicx}

\begin{document}

\title{High resolution characterisation of microstructural evolution in Rb$_x$Fe$_{2-y}$Se$_2$ crystals on annealing}

\author{S.C. Speller} 
\affiliation{Department of Materials, University of Oxford, OX1 3PH, UK}
\email{susannah.speller@materials.ox.ac.uk} 
\author{P. Dudin}
\affiliation{Diamond Light Source, Harwell Science and Innovation Campus, Didcot, OX11 0DE, UK}
\author{S. Fitzgerald}
\author{G.M. Hughes}
\author{K. Kruska}
\affiliation{Department of Materials, University of Oxford, OX1 3PH, UK}
\author{T.B. Britton} 
\affiliation{Department of Materials, Imperial College, London, SW7 2AZ, UK} 
\author{A. Krzton-Maziopa}
\affiliation{Warsaw University of Technology, Faculty of Chemistry, Noakowskiego St.3, 00-664 Warsaw, Poland}
\author{E. Pomjakushina}
\author{K. Conder}
\affiliation{Laboratory for Developments and Methods, Paul Scherrer Institut, CH-5232 Villigen PSI, Switzerland }
\author{A. Barinov}
\affiliation{Elettra-Sincrotrone Trieste ScPA, 34149 Basovizza, Trieste,Italy}
\author{C.R.M. Grovenor}
\affiliation{Department of Materials, University of Oxford, OX1 3PH, UK}

\begin{abstract}
The superconducting and magnetic properties of phase-separated A$_x$Fe$_{2-y}$Se$_2$ compounds are known to depend on post-growth heat treatments and cooling profiles.  This paper focusses on the evolution of microstructure on annealing, and how this influences the superconducting properties of \RFS\ single crystals.  We find that the minority phase in the as-grown crystal has increased unit cell anisotropy (\ca\ ratio), reduced Rb content and increased Fe content compared to the matrix.  The microstructure is rather complex, with two-phase mesoscopic plate-shaped features aligned along $\{113\}$ habit planes.  The minority phase are  strongly facetted on the $\{113\}$ planes, which we have shown to be driven by minimising the volume strain energy introduced as a result of the phase transformation.  Annealing at 488K results in coarsening of the mesoscopic plate-shaped features and the formation of a third distinct phase.  The subtle differences in structure and chemistry of the minority phase(s) in the crystals are thought to be responsible for changes in the superconducting transition temperature.   In addition, scanning photoemission microscopy has clearly shown that the electronic structure of the minority phase has a higher occupied density of states of the low binding energy Fe3d orbitals, characteristic of crystals that exhibit superconductivity.  This demonstrates a clear correlation between the Fe-vacancy-free phase with high \ca\ ratio and the electronic structure characteristics of the superconducting phase.  
\end{abstract}

\maketitle

%\section*{Introduction}
\section{Introduction}
Binary FeSe, with a superconducting transition temperature of 8K, is the simplest of the family of iron-based superconductors, consisting of tetrahedrally co-ordinated FeSe layers stacked with no spacing atoms \cite{Fang:2008}. Its transition temperature can be increased to 14K by  substitution of about half of the Se atoms with larger Te atoms \cite{Yeh:2008}. In 2010 it was discovered that the introduction of potassium atoms between the FeSe layers in the crystal structure produced a ternary compound with nominal composition of K$_x$Fe$_2$Se$_2$, and significantly increases the superconducting transition temperature to $\approx$ 30K  \cite{Guo:2010}.  Subsequently superconductivity has been found in a range of compounds in this family (A$_{x}$Fe$_{2-y}$Se$_2$ where A=K, Cs, Rb, Tl, etc.). The compositions of these compounds are well-known to deviate from the ideal stoichiometry \cite{Mou:2011}, with Fe-vacancies introduced  into the structure owing to restrictions on the valency of the iron atom.  At least five different types of iron ordering have been found in A$_{x}$Fe$_{2-y}$Se$_2$ compounds using both bulk techniques such as X-ray  and neutron diffraction and high resolution techniques such as TEM and STM microscopy, as discussed in a review article by Mou et al. \cite{Mou:2011}
Experimentally determined phase diagrams  for Rb$_x$Fe$_{2-y}$Se$_2$ \cite{Tsurkan:2011} and K$_x$Fe$_{2-y}$Se$_2$ \cite{Yan:2012} indicate that antiferromagnetic (AFM) ordering occurs at temperatures above 500K, with superconductivity co-existing with this AFM phase over a  narrow range of compositions.  Compounds with compositions either side of this region are insulating (or semiconducting), exhibiting different forms of AFM ordering and vacancy ordering schemes.  There is an ongoing debate about the parent compound of this system.   Some suggest it is the insulating $\sqrt5$x$\sqrt5$ ordered Fe vacancy phase with composition A$_{0.8}$Fe$_{1.6}$Se$_{2}$ (known as the ``245'' phase) which exhibits block AFM ordering \cite{Taylor:2012, Fang:2011, Yan:2012}.  In order to obtain the superconducting phase, extra Fe must be added as Fe vacancies are considered to be detrimental to superconductivity \cite{Li:2011, Texier:2012}.  Alternatively there has been speculation that the parent compound is a semiconducting phase (based on ARPES experiments) \cite{Chen:2011} or an $\sqrt8$x$\sqrt10$  ordered Fe vacancy phase with composition A$_{2}$Fe$_{7}$Se$_{8}$ (based on chemical microanalysis and STM) \cite{Ding:2013}.  

SEM \cite{Speller:2012, Ryan:2011, Liu:2012}, TEM \cite{Li:2011, Wang:2011}, STM \cite{Sun:2014}, nanofocused XRD studies \cite{Ricci:2011b} and SNOM \cite{Charnukha:2012} have all shown that strong phase separation exists in crystals that exhibit large shielding fractions in magnetisation measurements, and this two phase nature is supported by muon-spin rotation \cite{Shermadini:2012, Wang:2012}, M\"{o}ssbauer\cite{Ryan:2011}, X-ray and neutron diffraction \cite{Pomjakushin:2012} and  NMR \cite{Texier:2012} experiments, all indicating that the volume fraction of superconducting phase is small ($\approx$10\%) even in crytals exhibiting 100\% magnetic shielding. The composition of the minority superconducting phase is still under debate.  Whilst there is consensus about the low concentration of Fe vacancies, the alkali metal content is less clear, with NMR results reporting a composition of Rb$_{0.3}$Fe$_{2}$Se$_{2}$ \cite{Texier:2012} whereas refinement of neutron powder diffraction patterns gives  Rb$_{0.6}$Fe$_2$Se$_2$ \cite{Pomjakushin:2012} and high quality plan view TEM/EDX on potassium compounds gives a composition of K$_{0.5}$Fe$_{2}$Se$_{2}$ \cite{Sun:2014}.  It is generally believed that this minority phase is superconducting, whilst the matrix is the insulating "245" vacancy ordered phase.  The morphology of the minority phase in Cs compounds is shown by cross-sectional HR-SEM to consist of a three dimensional network of interconnected plates on the meso-scale \cite{Speller:2012} which explains some of the apparently contradictory properties of these compounds, such as high \Tc\ values and large shielding fractions coupled with large antiferromagnetic volume fractions and high normal state electrical resistivity.  

Post-growth heat treatments  have been found to strongly influence the superconductivity in \RFS\ crystals, with significant improvements to onset \Tc\ and transition width achieved with annealing and quenching from modest temperatures.   The cooling rate is also found to be crucial to the superconducting properties, with fast cooling from the growth temperature required to achieve high shielding fractions \cite{Liu:2012, Zhang:2013}.  This paper addresses the effects of annealing on microstructural development in the \RFS\ system in order to understand the influence of microstructure on the superconducting and magnetic properties, and to enable optimisation of the  growth process.  

%methods
\section{Experimental methods}
\subsection{Samples}
 Single crystals of \RFS\ were prepared by Bridgman growth using the process described elsewhere \cite{Krzton-Maziopa:2012}.   The as-grown (AG) crystal has been cooled from 750K  by quenching at a rate of -200K/min.  The crystal was then separated into different fragments in a He glovebox and resealed under vacuum in separate quartz tubes to prevent chemical degradation.  Two samples were then annealed for 3 hours at 488K (A488) and 563K (A563), followed by quenching.   These temperatures were selected as they correspond to the phase separation temperature and a temperature above the ordering temperature respectively \cite{Weyeneth:2012, Pomjakushin:2012}.  A wide range of  superconducting and magnetic property measurements including magnetisation, transport resistivity and  muon spin rotation ($\mu$SR) have previously been reported for earlier batches of crystals grown and heat treated under identical conditions \cite{Weyeneth:2012}, and structural data from X-ray and neutron diffraction have also been obtained on AG crystals before and after annealing at 100 hours at 488K \cite{Pomjakushin:2012}.  All samples exhibited 100\% magnetic shielding in magnetization measurements, and annealing at 488K and quenching resulted in an improvement in the superconducting properties, with slightly increased \Tc\ values (of $\approx$ 5\%) and significantly decreased transition widths \cite{Weyeneth:2012}.  Annealing at the higher temperature of 563K was found to be detrimental, reducing the onset \Tc\ to about 20K.  Both $\mu$SR and neutron diffraction confirmed the presence of $\approx$10\% minority (superconducting) phase in the AG crystals.  The crystal structure of the minority phase is found to be compressed along the a-axis and expanded along the c-axis. However the $\mu$sr results indicate that the volume fraction of superconducting (non-magnetic) phase was unchanged upon annealing for 60 hours at 488K, but the non-magnetic phase is found to be distributed on a finer spatial scale after annealing.   In contrast, the neutron experiments showed a decrease in the volume fraction of the minority superconducting phase for a crystal annealed at the same temperature for the longer time of 100 hours.  

\subsection{Microstructural Characterisation}
High-resolution electron backscatter diffraction (HR-EBSD) experiments were carried out on freshly cleaved (001) surfaces in a JEOL 6500F SEM using an EDAX/TSL Digiview II detector.  This technique is used to map changes in unit cell anisotropy, through cross correlation of electron diffraction patterns and subsequent analysis (described in \cite{Speller:2011}). HR-EBSD can determine relative changes in c and a values, but cannot determine absolute c and a values.  Therefore the absolute c/a values have been calculated by setting the modal c/a ratio of each map to 3.711 (based upon XRD measurements \cite{Pomjakushin:2012}). Cross-sectional imaging and serial sectioning were performed on a Zeiss NVision FIB, with Avizo software used for 3D image reconstruction.  A JEOL 5510 SEM with an Oxford Instruments X-act EDX detector was used for SEM and quantitative EDX and a Zeiss Merlin SEM with a 150mm$^2$ Oxford Instruments X-max detector was used for low voltage SEM imaging and high resolution EDX.  Scanning photoemission microscopy (SPEM) was performed at the SpectroMicroscopy beamline at Elettra synchrotron \cite{Dudin:2010} using a 74eV photon light source. 

%results
\section{High resolution electron backscatter diffraction analysis}
 Figure \ref{fig:maps} shows typical HR-EBSD maps from AG, A488 and A563 samples.  The microstructure of the AG sample (fig. \ref{fig:maps} a) is very similar to the \CFS\ samples studied previously \cite{Speller:2012}, consisting of a square network of linear features oriented along the \textless110\textgreater\ directions with a larger \ca\ ratio than the matrix. The three-dimensional morphology associated with these features can be seen from cross-sectional FIB/SEM to consist of a network of plates aligned along the \{113\} habit planes, as previously seen in both the Cs  and K analogues \cite{Speller:2012, Wang:2014}.  The plates typically appear  discontinuous (stripey) as more clearly seen in the 3D reconstruction of the minority phase produced by serial FIB polishing in cross-section (fig. \ref{fig:3DFIB}).
% A recent STEM study on the as-grown \RFS\ crystals has also shown that the main linear features break down into a finger-shaped morphology, giving rise to the stripey appearance in 2D sections \cite{Weyeneth:2012}. 

Annealing at the nominal phase separation temperature (488K) results in two main changes to the microstructure; the separation of the mesoscopic plate features increases from $\approx$ 3 $\mu$m to 5-10 $\mu$m, and the \ca\ ratio within these features decreases to a value closer to that of the matrix phase.  The mesoscopic features are also less regularly  spaced with  a larger variability in their \ca\ values.  The EBSD pattern quality is found to be considerably poorer within the plate-shaped features, probably owing to even finer scale phase separation (stripes) within the plates.  This makes it possible to isolate the minority phase by thresholding the maps using the mean angular error (MAE) parameter from the image correlation procedure (see figs. \ref{fig:maps}d and e).  Whilst it is not possible to obtain an accurate measure of the volume fraction of minority phase from this process as the finer scale phase separation is not resolved, it is clear that in the AG sample all the plate features have higher \ca\ ratio than the matrix.  However, in the sample annealed at 488K, some of the plates have higher \ca\ ratio but others have similar or lower \ca\ ratio than the matrix, suggesting the presence of at least two different phases in addition to the matrix.  
Annealing at the higher temperature of 563K, above the Fe vacancy ordering temperature of $\approx$ 541K, produces a much more homogeneous microstructure on the length scale probed by the HR-EBSD technique (\textgreater 100nm) with only small fluctuations of $\Delta$\ca \textless 0.005 on the length scale of $\approx$ 1$\mu$m and with no sign of the mesoscopic plate-shaped features seen in the AG and A488 samples.  Similar matrix fluctuations are observed in the matrix of the A488 samples and are within the noise threshold. 

%The EBSD results indicate the mean increase in \ca\ ratio of the plate features in the AG sample relative to the majority phase is $\Delta$\ca \textless 0.5\%, whereas Pomjakushin et al. calculate from neutron scattering experiments a significantly larger shift of $\Delta$\ca $\approx$ 3.5\% for the minority Rb$_x$Fe$_2$Se$_2$ phase \cite{Pomjakushin:2012}. This is believed to be a result of the nano-scale stripey phase separation within the plate-shaped features.  The  EBSD patterns would be a superposition of the patterns for the high \ca\ phase with a lower \ca\ phase, which in practice produces a smearing of the  bands on the diffraction pattern, resulting in an intermediate value of \ca\ being measured. 

The probability distribution of \ca\ values for the AG sample shows a skew towards higher \ca\ values with a shoulder visible on the high \ca\ side of the peak.  This distribution has been fitted to a two Gaussian model, with the majority peak having  mean \ca=3.711, $\sigma$=0.0046 and the minority peak mean \ca=3.721 $\sigma$=0.0097.  This increase in \ca\ ratio in the minority phase relative to the matrix is considerably smaller than the 3.5\% increase found by XRD on similar Rb crystals \cite{Pomjakushin:2012}. This is believed to be a result of the nano-scale stripey phase separation within the plate-shaped features. If the volume from which the backscattered electrons originated in the sample is bigger than the secondary phase domains, the  EBSD patterns would be a superposition of the patterns for the high \ca\ phase with the lower \ca\ matrix, which in practice produces a smearing of the  bands on the diffraction pattern, resulting in an intermediate value of \ca\ being measured. The Gaussian fits for the majority peaks in both the AG and A488 samples have $\frac{\textrm{FWHM}}{\textrm{average} (c/a)}$ of 0.3\%, identical to the  neutron scattering data \cite{Pomjakushin:2012}.  The \ca\ distribution of the A563 sample is well described by a single Gaussian distribution with the smaller standard deviation of 0.0035, indicating a reduction in the extent of chemical inhomogeneity and/or local mechanical strains in this sample.  

%EDX and BSE
\section{Chemical composition analysis}
The multi-phase microstructures present in the AG and A488 samples have been studied further in the SEM using backscattered electron (BSE) imaging in compositional mode and EDX microanalysis (fig. \ref{fig:5510} and \ref{fig:merlin}).  The increase in the length-scale of the mesoscopic plate-shaped features on annealing  is immediately apparent in fig. \ref{fig:5510}, and thresholding the BSE image contrast using ImageJ software gives an areal fraction of the dark contrast phase of approximately 10\% in both samples.  If we correlate these plate-shaped features to the superconducting phase, this value is consistent with neutron scattering data  \cite{Pomjakushin:2012} and muon-spin rotation studies \cite{Weyeneth:2012}.  The improved spatial resolution relative to the sampling volume used in the HR-EBSD maps shows that the spacing of the fine-scale stripey phase separation within the plates also tends to increase on annealing at 488K.    In the AG sample, BSE images suggest that the mesoscopic plates consist of two phases: one with the same contrast as the matrix and the other appearing darker (lower average atomic number).  Interestingly, in contrast to the AG samples, three distinct BSE contrasts can be seen in the annealed sample, suggesting the presence of two different minority phases within the matrix, one with darker BSE contrast and the other with brighter BSE contrast.  Quantitative EDX at 20kV gave a matrix composition of Rb$_{0.81(1)}$Fe$_{1.55(2)}$Se$_{2}$ and Rb$_{0.81(1)}$Fe$_{1.58(2)}$Se$_{2}$ for the AG and A488 samples respectively, consistent with the matrix being the AFM insulating ``245'' phase.  The precise composition of the minor phases cannot be measured directly as the interaction volume at 20kV (required to fully excite the K Se line for quantitiative analysis) is significantly larger than the minority phase domains.  However,  since the overall composition measured over a large area of the crystal lies on a line between the matrix composition and the measured composition of the minority phase, by estimating the minority phase volume fraction of 10\% we find the actual minority phase composition to be approximately Rb$_{0.5}$Fe$_{2}$Se$_{2}$, consistent with TEM analysis fo minority phases in the potassium compounds \cite{Sun:2014}.

The three-phase microstructure of the 488K annealed crystal was further investigated using 5kV accelerating voltage to significantly reduce the EDX interaction volume.  Figure \ref{fig:merlin} clearly shows the plates with dark BSE contrast are Fe-rich and Rb-deficient compared to the matrix as expected.  Some of the mesoscopic plates consist of two phases as seen in fig. \ref{fig:merlin}(d)-(f) where one phase appears to be the same composition as the matrix and the other is the Fe-rich, Rb-deficient phase with dark BSE contrast.  However, in other regions there are some plates that have brighter contrast than the matrix and this bright phase is sometimes seen as one of the phases within stripey features as clearly seen in fig. \ref{fig:merlin}(g)-(i).  By comparing the spectra (not shown) from the regions indicated in fig. \ref{fig:merlin}(g)  it can be seen that the bright phase has an intermediate composition, still Fe-rich but less Rb deficient than the dark phase.  This alkali metal partitioning on the nano-scale has also been reported recently for the K compounds and has been attributed to K-ordering within one of the phases \cite{Wang:2014}.

We conclude that the plate-shaped features consist of two phases in the as-grown crystal: the ordered Fe vacancy phase with composition close to Rb$_{0.8}$Fe$_{1.6}$Se$_2$ and the "vacancy-free" superconducting \RFS\ phase with y$\approx$0 and increased \ca\ ratio relative to the matrix.  In the annealed crystals there are at least two different minority phases present in addition to the matrix: the dark contrast vacancy-free phase (probably with increased \ca\ ratio) and a bright contrast phase that is also Fe-rich but less Rb deficient than the dark vacancy free phase that may be the lower \ca\ ratio phase seen in the HR-EBSD map.   It is interesting to note that HAADF STEM analysis on \KFS\ crystals has also identified another phase within the Fe-rich plates which has a lower c-axis lattice parameter and containing very fine scale low atomic number features in the (001) plane that may be cracks formed to relieve residual stress in the multiphase system.  These very fine scale features have also been observed by other researchers \cite{Weyeneth:2012, Charnukha:2012}. 

\subsection*{Microstructural development}
In order to understand the effects of heat treatments on the superconducting and magnetic properties, it is beneficial to discuss the thermodynamic and kinetic factors affecting the complex microstructural development in this system.  Previously reported DSC measurements on  earlier batches of \RFS\ crystals \cite{Pomjakushin:2012, Weyeneth:2012} show peaks corresponding to the onset of the ordering of Fe vacancies at T$_S \approx$ 540K, the AFM Neel temperature T$_N \approx$ 517K and a third peak attributed to the phase separation into the ordered vacancy phase plus the vacancy-free phase at T$_P \approx$ 488K. The growth process involves  cooling crystals at a moderate rate from the disordered Fe vacancy phase at high temperature.  On cooling through T$_S$, the ordered Fe vacancy phase forms, probably by a nucleation and growth process, rejecting excess Fe ahead of the growth interface.  Wang et al. suggest from their in-situ TEM studies that the Fe-vacancy ordering occurs by a spinodal mechanism, but no conclusive evidence for this has been presented.  The length scale of the ordered domains are very large for spinodal microstructures, especially at low temperatures, and the interfaces between the domains appear abrupt which is more indicative of a nucleation and growth process than spinodal decomposition.  The enriched Fe domains that remain between the ordered vacancy phase domains eventually decompose into two (or more) phases at T$_P$ producing the fine stripey appearance clearly seen in figs. \ref{fig:merlin}(d) and (g).  It is possible that there are a number of binary eutectoid reactions in system and slight fluctuations in the composition of the un-transformed Fe-rich phase result in different volume fractions of the three phases being produced on cooling.  The scale of the mesoscopic plate-like microstructure is governed by the rate of diffusion that can occur below T$_S$ during processing.  The moderately fast cooling in the as-grown crystal results, as expected, in smaller scale plate separation than the samples annealed at temperatures below T$_S$ (e.g. 488K) in which coarsening of the mesoscopic Fe-rich plates occurs.   After annealing, the quench is expected to be faster than the as-grown cooling rate, as the size of the fragments of the crystals annealed is considerably smaller than the initial crystals.  This explains why the sample annealed at 563K (above T$_S$) has a microstructure that is too fine to be seen in the HR-EBSD.  The sample annealed at 488K has slightly coarser phase separation within the plates than the AG sample, suggesting that the annealing temperature is slightly below T$_P$.  However, the muon spin rotation data on a crystal prepared under similar annealing conditions finds a reduced size of the paramagnetic (superconducting) domains.  In this case it is likely that the annealing temperature is slightly above T$_P$ and the fine scale microstructure is produced on quenching through T$_P$.   Very small differences in crystal composition or furnace temperature could be responsible for the differences between the sample studied here and the one  used in the muon spin rotation experiment.  

%electronic properties
\section{Electronic properties}
It remains to explore the reason why annealing at 488K improves T$_c$ of the superconducting phase.  One possibility is that the chemistry and resulting structural parameters of the superconducting phase change slightly as a result of the annealing and quenching procedure.  This could result from local equilibriation of the composition of the Fe-rich phase above T$_P$, different cooling rate affecting the composition and/or strain in the superconducting phase.  Alternatively it could be  the finer scale of the phase separation within the mesoscopic plates that is responsible for the improved superconducting properties, with suggestions that superconductivity in this system is strongly influenced by the surrounding AFM matrix phase, with proximity effects enhancing T$_c$ \cite{Huang:2013}.  Therefore, the finer scale microstructures within the plates seen in muon-spin studies on crystals annealed at 488K \cite{Weyeneth:2012}  might be responsible for the improved properties.  

To investigate the electronic properties of the different phases, scanning photoemission microscopy (SPEM) has been performed using the SpectroMicroscopy beamline at Elettra.  Figure \ref{fig:spectromicroscopy} compares an as-grown crystal (AG-2) with one annealed at 488K for 3 hours (A488-2).  These samples are from a different growth run than the crystals studied above, but they have been prepared under the same conditions.
The maps in figure \ref{fig:spectromicroscopy} (a) and (d) show the spatial variation in the intensity of photoelectrons emitted within the energy window between 71 and 69.7 eV (corresponding to low binding energies up to -1.15eV relative to the Fermi level at 70.85 eV) for the AG-2 and A488-2 samples respectively.  The mesoscopic plate-like microstructure is clearly visible as bright features in these near-Fermi-level maps, with the scale of the microstructure being larger for the annealed sample as observed in the SEM.   The lateral dimensions of these mesoscopic plate features appears to be larger in the SPEM images than in the SEM, possibly because SPEM is a much more surface sensitive technique.   The fine-scale phase separation within the plates is not typically resolved in the SPEM images which may be a result of the inherent lower lateral spatial resolution (above 0.6$\mu$m), or it may be an indication that the electronic structure in the matrix is modified in the vicinity of the secondary phase. The survey spectra in \ref{fig:spectromicroscopy} (b) and (e) show that the main difference between the bright and dark regions in the maps is the intensity of the lowest binding energy peaks which are known from DFT modelling  to be mainly associated with Fe 3d orbitals \cite{Yan:2011, Ivanovskii:2011}.  The more detailed valence band spectra in figures \ref{fig:spectromicroscopy} (c) and (f) reveal that the  low binding energy feature consists of at least two distinct peaks, with the intensity of the peak just below the Fermi level substantially higher in the plate features relative to the matrix phase.  Comparing the spectra from bright regions in the AG and A488 samples shows that the increased contrast seen in the A488 image results from a significant enhancement in the lowest binding energy peak.  Using classical ARPES  Chen et al. found that this peak originates from the superconducting phase in \KFS\ samples \cite{Chen:2011}, providing direct evidence that the bright plate-like  features seen in our SPEM images are the phase that becomes superconducting at lower temperatures, with the dark matrix phase having insulating properties.  A simple argument based on the relative number of valence electrons per unit cell in the vacancy-ordered Rb$_{0.8}$Fe$_{1.6}$Se$_{2}$ phase (25.6 e per formula unit) and the vacancy-free composition Rb$_{0.5}$Fe$_{2}$Se$_{2}$ phase (28.5 e per formula unit) suggests that the higher Fe content in the vacancy free phase leads to increased occupation of the upper Fe 3d orbitals leading to metallic/superconducting properties.  More detailed analysis of the SPEM results will be the topic of a future publication.

\section{Elastic properties}

The $\{113\}$ habit planes of the facetted minority phase domains can be understood through the stress-free transformation strain required to create them from the parent phase. Assuming a coherent interface, this is given by 
\begin{equation}
\bm{\epsilon^0} = \frac{F+F^T}{2} - I, \;\; {\rm where}\;\; F = \left(\begin{array}{ccc}
a_2/a_1 & 0 & 0\\
0 & a_2/a_1 & 0\\
0 & 0 & c_2/c_1\end{array}\right),
\end{equation} and $a_1,c_1,a_2,c_2$ are the lattice parameters of the parent and minority phases respectively. According to the Khachaturyan-Shaltov microelasticity theory, the precipitates will form facets with normal vectors $\bm{n}$ that minimize the following energy function: 
\begin{equation}
%B(\bm{n}) = \bm{\sigma^0\cdot\epsilon^0} - \bm{n\cdot\sigma\cdot\Omega\cdot\sigma\cdot n}
B(\bm{n}) = \sigma^0_{ij}\epsilon^0_{ij} - n_i\sigma^0_{ij}\Omega_{jk}(\bm{n})\sigma^0_{kl}n_l, 
\end{equation} where $\sigma^0_{ij}=c_{ijkl}\epsilon^0_{kl}$ and $\Omega_{jk}(\bm{n})=(n_ic_{ijkl}n_l)^{-1}$ (see Shi {\it et al}\cite{Shi:2012}). Taking the elastic moduli calculated for a similar compound \cite{Ivanovskii:2011}, and the lattice parameters observed at the phase separation temperature \cite{Pomjakushin:2012}, gives the energy function $B(\bm{n})$ shown in figure \ref{fig:elasticity}. The left pane shows the full three-dimensional energy surface, and the centre pane shows its projection onto the $(1\bar 1 0)$ plane for ease of comparison with figure \ref{fig:3DFIB}. This clearly shows that the directions corresponding to the minimum $B$ values are at an angle of 30$^{\circ}$ to [110] direction, along the $\{113\}$ plane normal directions.  This explains that the facetting we observe on the $\{113\}$ planes as shown in the right pane of figure \ref{fig:elasticity} arises as a result of minimising the volume strain energy associated with the transformation.

%Conclusion
\section{Conclusion}
In conclusion, we have investigated spatial variations in crystal structure, chemistry and electronic structure in \RFS\ crystals subjected to three different thermal treatments using HR-EBSD, SEM microanalysis and SPEM.  By combining our microstructural results with earlier structural and magnetic/superconducting measurements, we have developed an explanation of the phase transformations occuring in this system which is in agreement with the physical properties. The mesoscopic plate-shaped features are believed to develop upon Fe vacancy ordering, originating from the residual high Fe content disordered vacancy phase left between the facetted domains of the ordered phase.   We have shown that the observed facetting on the $\{113\}$ planes can be explained from calculations of the elastic strain energy associated with the transformation.  The metastable microstructures formed during moderately fast cooling from high temperatures can be modified substantially by subsequent annealing and quenching. The HR-EBSD technique cannot image the nanoscale phase separation in these crystals, but the pattern quality deterioration in the mesoscopic features is indicative of the presence of multiple phases at the nanoscale.  The SPEM results are consistent with the minority high Fe-content, high \ca\ phase being superconducting and the lower \ca\ matrix having insulating properties.   Further microstrucural studies using higher spatial resolution techniques such as (S)TEM with vacuum transfer capability to avoid reaction with air are necessary for quantitative characterisation of the nano-scale phase separation in these materials.

%\begin{acknowledgements}
Dr. S. Speller was supported under the RAEng/EPSRC research fellowship scheme.  Dr. A. Krzton-Maziopa acknowledges the financial support founded by the National Science Centre of Poland grant No. DEC-2013/09/B/ST5/03391.  

%\end{acknowledgements}

%Bibliography
%\section*{References}
\bibliographystyle{unsrt}

%figures
\begin{figure}[h]
  \begin{center}
 \end{center}
  \caption{HR-EBSD maps from (a) as-grown, (b)488K annealed and (c) 563K annealed crystals showing spatial variations in \ca\ ratio.  HR-EBSD maps for AG (d) and A488 (e) thresholded on MAE\textgreater 0.009 and MAE\textgreater 0.007 respectively.  (f) gives the probablilty distribution of c/a ratio for each map, with the white pixels (poor quality data) removed.}
\label{fig:maps}
\end{figure}

\begin{figure}[h]
  \begin{center}
 \end{center}
  \caption{Three dimensional reconstruction of the minority phase in the as-grown crystal produced by serial FIB sectioning. The yellow phase near the surface of the crystal is oxide which is present as the surface was not freshly cleaved for the cross-sectional analysis. The reconstructed volume is 10x7.5x0.8$\mu$m in size.}
\label{fig:3DFIB}
\end{figure}

\begin{figure}[h]
  \begin{center}
 \end{center}
  \caption{20kV backscattered electron images showing compositional variations in AG (left pane) and (b) A488 samples (right pane).}
\label{fig:5510}
\end{figure}

\begin{figure}[h]
  \begin{center}
 \end{center}
  \caption{BSE images with corresponding EDX maps taken at 5kV for three different regions of A488 sample.}
\label{fig:merlin}
\end{figure}

\begin{figure}[h] 
  \begin{center}
 \end{center}
  \caption{Scanning photoemission microscopy data for samples AG-2 (a-c) and A488-2 (d-f): near-Fermi-level maps (a) and (d), survey spectra (b) and (e) and detailed spectra near the Fermi level (c) and (f).}
\label{fig:spectromicroscopy}
\end{figure}

\begin{figure}[h] 
  \begin{center}

 \end{center}
  \caption{Left: Calculated energy surface as a function of facet normal $\bm{n}$. Centre: projection onto $(1\bar 1 0)$ plane (arbitrary units).  Right: SEM micrograph showing the facetting in the  $(1\bar 1 0)$ plane}
\label{fig:elasticity}
\end{figure}

%\section*{References}
%\bibliographystyle{unsrt}
%\bibliography{CFS}
\end{document}